\documentclass[aps,prb,twocolumn,showpacs,amsmath,amssymb,superscriptaddress,floatfix]{revtex4-1}
\usepackage{graphicx}
\usepackage{dcolumn}
\usepackage{bm}
\usepackage{epsfig}
\usepackage{epstopdf}
\usepackage{float}	
\usepackage{color}
\include{graphic}

\begin{document}
\title{Cavity mediated dissipative coupling of distant magnetic moments: theory and experiment}

\author{Peng-Chao~Xu} \email{pcxu14@fudan.edu.cn;}
\affiliation{State Key Laboratory of Surface Physics and Department of Physics, Fudan University, Shanghai 200433, China}
\affiliation{Collaborative Innovation Center of Advanced Microstructures, Fudan University, Shanghai 200433, China}
\affiliation{Department of Physics and Astronomy, University of Manitoba, Winnipeg, Canada R3T 2N2}
\author{J. W.~Rao}
\affiliation{Department of Physics and Astronomy, University of Manitoba, Winnipeg, Canada R3T 2N2}
\author{Y. S.~Gui}
\affiliation{Department of Physics and Astronomy, University of Manitoba, Winnipeg, Canada R3T 2N2}
\author{Xiaofeng~Jin}
\affiliation{State Key Laboratory of Surface Physics and Department of Physics, Fudan University, Shanghai 200433, China}
\affiliation{Collaborative Innovation Center of Advanced Microstructures, Fudan University, Shanghai 200433, China}
\author{C.-M.~Hu} \email{Can-Ming.Hu@umanitoba.ca;}
\affiliation{Department of Physics and Astronomy, University of Manitoba, Winnipeg, Canada R3T 2N2}

\begin{abstract}
We investigate long-range coherent and dissipative coupling between two spatially separated magnets while both are coupled to a microwave cavity. A careful examination of the system shows that the indirect interaction between two magnon modes is dependent on their individual mechanisms of direct coupling to the cavity. If both magnon modes share the same form of coupling to the cavity (either coherent or dissipative), then the indirect coupling between them will produce level repulsion. Conversely, if the magnon modes have different forms of coupling to the cavity (one coherent and one dissipative), then their indirect coupling will produce level attraction. We further demonstrate the cavity-mediate nature of the indirect interaction through investigating the dependence of the indirect coupling strength on the frequency detuning between the magnon and cavity modes. Our work theoretically and experimentally explores indirect cavity mediate interactions in systems exhibiting both coherent and dissipative coupling, which opens a new avenue for controlling and utilizing light-matter interactions.

\end{abstract}

\maketitle

\section{introduction}
Light matter interactions in a hybrid system are of great interest in modern physics as a building block for coherent information processing.\cite{Wallraff2004Nature, Kubo2010PRL, Putz2014NP, Wallquist2009physscr} Ideally, two distant quantum systems can transfer information through their mutual coupling to a resonant photon system, which is highly desirable for any architecture of quantum information processing.\cite{Majer2007Nature, Xiang2013RMP, Kurizki2015PNAS, Dany2019APE} As a newly discovered form of light-matter interaction, the strong coupling between microwave photons and magnons has been attracting increasing attention in recent years, since the interaction between electrodynamics and magnetization dynamics spawns the cavity magnon polariton (CMP).\cite{Soykal2010PRL, Huebl2013PRL, Zhang2014PRL, Tabuchi2014PRL, Goryachev2014PRA, Cao2015PRB, Bai2015PRL, Yao2015PRB, Lambert2015JAP} The CMP dispersion manifests as an elegant level repulsion, in which lies the profound physics of Rabi splitting. \cite{Khitrova2006NP} The high spin density and low room temperature damping rate of ferromagnetic insulators allows the coherent photon-spin interactions to enter the ultrastrong coupling regime. \cite{Zhang2014PRL, Goryachev2014PRA, Niemczyk2010NP} Consequently, the magnon photon coupling provides a perfect platform for studying and illustrating coupling related physics, specifically, in addition to being a promising candidate for coherent information processing. \cite{Dany2019APE, Rao2019NC}

Interestingly, when a cavity mode couples to two macroscopic magnetic moments simultaneously, the quantized magnetic field of the photons links the dynamics of distant magnons, thus inducing nonlocal interaction between the magnon modes. \cite{Haigh2015PRB, Lambert2016PRA, Bai2017PRL, Zare2018PRB} In turn, the coherent transport of magnons may survive over macroscopic distances. Recently, long range dispersive coupling between two macro-spin systems has been demonstrated experimentally by a model system with two ferrimagnets within a cavity. \cite{Haigh2015PRB, Lambert2016PRA, Zare2018PRB} Combined with state of art spintronics technology, nonlocal spin current manipulation over several centimeters has also been achieved.\cite{Bai2017PRL} In fact, photon mediated magnon coupling is so ubiquitous that it may even arise between ferromagnets and antiferromagnets.\cite{Johansen2018PRL} In this sense, cavity mediated coupling can combine ferromagnetic and antiferromagnetic spintronics within the frame of cavity spintronics. On the storage end, gradient memory architectures have been developed making use of the phase correlation and scalability of CMP systems.\cite{zhang2015NC}

Aside from coherent coupling, a form of dissipative coupling was revealed recently in CMP systems. \cite{Harder2018PRL, Grigoryan2018PRB, Bhoi2019PRB, Yang2019PRA, Rao2019NJP, Boventer2019} In contrast to the level repulsion rising from coherent coupling between resonant modes, dissipative coupling shows an exotic level attraction. These two coupling mechanisms give us an unprecedented degree of freedom to control photon mediated interaction. 

In this work, we revisit cavity mediated long range interaction between two magnets from the new perspective of controlling coupling mechanisms. By placing two yttrium iron garnet (YIG) spheres at different positions in a cavity, which correspond to coherent coupling or dissipative coupling, we experimentally show that the indirect coupling between the two magnon systems act as a XOR-like logic gate in the coupling phenomenon. Specifically, when both magnon modes share the same form of coupling to the cavity (either coherent or dissipative), then the indirect coupling between them will produce level repulsion. Conversely, if the magnon modes have different forms of coupling to the cavity (one coherent and one dissipative), then their indirect coupling will produce level attraction. Treating dissipative coupling on an equal footing with coherent coupling, we are now able to establish the correlation between local magnon-photon coupling and non-local magnon-magnon coupling. Our work reveals the cavity mediated dissipative coupling between distant magnetic moments, which opens a new avenue for controlling and utilizing light-matter interaction.

This paper is split into two main sections, which discuss the theoretical model and experimental results. In the theoretical model part, we first provide a brief comparison between coherent and dissipative magnon-photon coupling, particularly, the evolution of the eigenvector in the coupled system, which allows us to clearly distinguish the cases of level repulsion and attraction. Then we present the formula describing long-range coherent and dissipative indirect magnon-magnon interactions through their mutual coupling to a cavity photon mode. Finally, we present the implementation of our experimental set-up and quantitatively compare the experimental observations with the theoretical model.

\section{THEORETICAL MODEL}
For a quantitative understanding of the long-range magnon-magnon interaction, we first study a general theoretical model involving coherent and dissipative coupling in a two-mode system. From this we clearly see the distinguishing features of level repulsion and level attraction in the dispersion, as well as the eigenvector of the coupled system. Then we consider the long-range indirect interaction between two magnon modes, which are coupled to a common microwave cavity. We rewrite the eigenfrequencies and eigenvectors of the system in a more explicit form using the dispersive approximation, where the frequency detuning between the magnon mode and the cavity mode is assumed to be large compare to the coupling strength. \cite{Schuster2007Nature} 

\subsection{Eigenfrequency and eigenvector in a strongly coupled magnon-photon system}
\label{Sec:2x2model}

We start with a general theoretical model of a two-mode system involving both coherent and dissipative coupling, which can be described by an equivalent non-Hermitian Hamiltonian\cite{Harder2018PRL} as

\begin{equation}
H=\hbar\omega_ca^\dag a+\hbar\omega_m m^\dag m+\hbar g(a^\dag m+e^{i\Phi}am^\dag),
\label{Eq:2x2Hamiltonian}
\end{equation}

\noindent where $\omega_c$ ($\omega_m$) is the frequency of the cavity (magnon) mode, $a^\dag$ ($a$) and $m^\dag$ ($m$) are the creation (annihilation) operators for the cavity and magnon mode respectively. The coupling rate $g$ is chosen to be a real positive number. The coupling phase $\Phi$ describes the competing coherent and dissipative couplings: $\Phi=0$ for level repulsion and $\Phi=\pi$ for level attraction. \cite{Harder2018PRL}

In the Heisenberg picture, this leads to the equation of motion 

\begin{equation}
\frac{d}{dt} \begin{pmatrix} a\\ m \end{pmatrix}
=i\begin{pmatrix}  \omega_c & g \\ e^{i\Phi}g & \omega_m \end{pmatrix}
\begin{pmatrix} a\\ m \end{pmatrix}.
\label{Eq:2x2ME}
\end{equation}

Following the $e^{-i\omega{}t}$ convention, The hybridized eigenmodes of the system are found by diagonalizing the matrix in Eq.(\ref{Eq:2x2ME}), and have eigenfrequencies

\begin{equation}
\omega_\pm=\frac{1}{2}\left[\omega_m+\omega_c\pm\sqrt{(\omega_m-\omega_c)^2+4e^{i\Phi}g^2}\right],
\label{Eq:2x2eigenfrequency}
\end{equation}

\noindent and eigenvectors
\begin{eqnarray}
\begin{pmatrix} a\\ m \end{pmatrix}
&=\begin{pmatrix}  g& \\
   \Delta_H/2\pm\sqrt{\Delta_H^2/4+e^{i\Phi}g^2}& \end{pmatrix}, \nonumber \\ 
&\mathrm{with}\qquad \Delta_H=\omega_m-\omega_c. 
\label{Eq:2x2eigenvector}
\end{eqnarray}

\begin{figure} [t!]
\begin{center}\
\epsfig{file=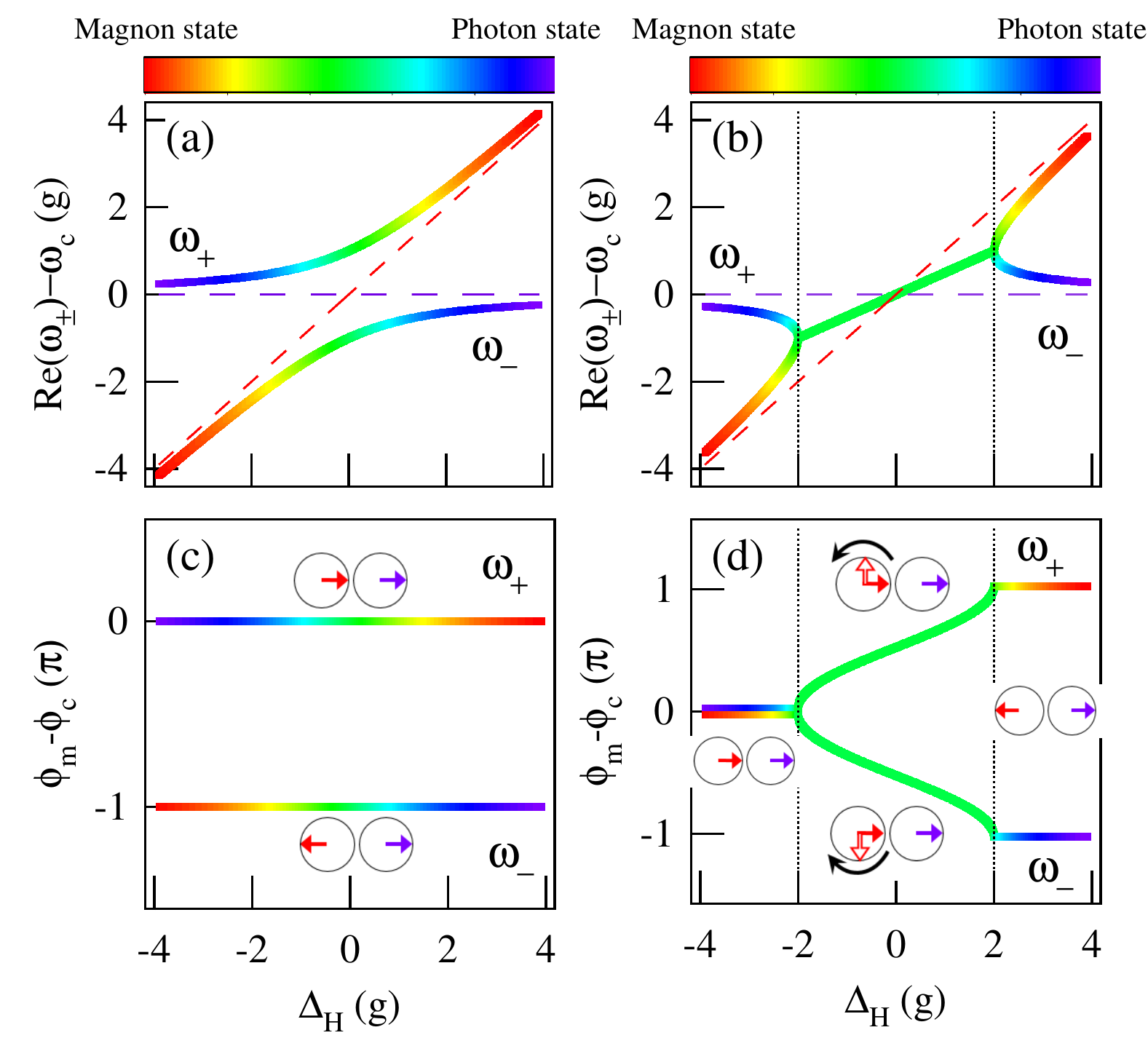,width=8 cm}
\caption{(a), (b) The hybridized mode frequencies, $\mathbf{Re}(\omega_\pm)-\omega_c$, are plotted as a function of the field detuning $\Delta_H=\omega_m-\omega_c$, for level repulsion and attraction calculated using Eq. (\ref{Eq:2x2eigenfrequency}) at $\Phi=0$ and $\Phi=\pi$, respectively. The red and blue dashed lines represent the uncoupled magnon mode $\omega_m$ and cavity mode $\omega_c$. The phase correlation, $\phi_m-\phi_c$, of magnetization and electrodynamics calculated by Eq. (\ref{Eq:2x2eigenvector}) are plotted as a function of $\Delta_H$ in (c) and (d), where inset figures show the phase difference between the magnon component $m$ (red) and photon component $a$ (blue). For simplicity, the phase of $a$ is set to be zero. Dotted lines in (b) and (d) indicated the condition of $\Delta_H=\pm2g$, where the two hybridized mode coalesce at the exceptional point. All curves are coloured by the contribution from $a$ (blue) and $m$ (red).}
\label{Fig1}
\end{center}
\end{figure}

In the frequency domain, two coupled modes (indicated by coloured lines) repel with each other for $\Phi=0$ and attract each other for $\Phi=\pi$ as clearly shown in Figs. 1(a) and (b) respectively. As a result, for level attraction, the two hybridized modes coalesce and have an identical eigenvector at a condition of $\Delta_H=\omega_m-\omega_c=\pm2g$, resenbling an exceptional point\cite{Heiss2012JPA, Zhang2017NC} through linear dynamics in the absence of damping. 

The phase correlation, $\phi_m-\phi_c$, of magnetization ($m$) and electrodynamics ($a$) calculated by Eq. (\ref{Eq:2x2eigenvector}) is shown in Figs. 1(c) and (d) for level repulsion and attraction, respectively. In level repulsion case, the phase correlation between $m$ and $a$ is quite simple:  $\phi_m-\phi_c=0$ (corresponding to in-phase $a\!-\!m$ motion) for $\omega_+$, and $\phi_m-\phi_c=-\pi$ (corresponding to out-of-phase $a\!-m\!$ motion) for $\omega_-$. For level attraction, it can be found that the phase correlation is completely different: for both $\omega_\pm$, $\phi_m-\phi_c=0$ when $\Delta_H<-2g$ and $\phi_m-\phi_c=\pi$ when $\Delta_H>2g$; in between those $\Delta_H$ values the $m$ phase for $\omega_+$ rotates anti-clockwise from 0 to $\pi$ with respect to $a$ while for $\omega_-$ the $m$ phase rotates clockwise from 0 to $-\pi$. Although the two hybridized magnon-photon modes follow an identical dispersion over a wide range for $|\Delta_H|<2g$ $\mathbf{Re}(\omega_\pm)=(\omega_c+\omega_m)/2$  and furthermore both consist of half magnon and half photon, the two states are independent because their correlation phases have opposite signs.

\subsection{Two distanced magnons coupled with a common cavity mode}

\begin{figure} [t!]
\begin{center}\
\epsfig{file=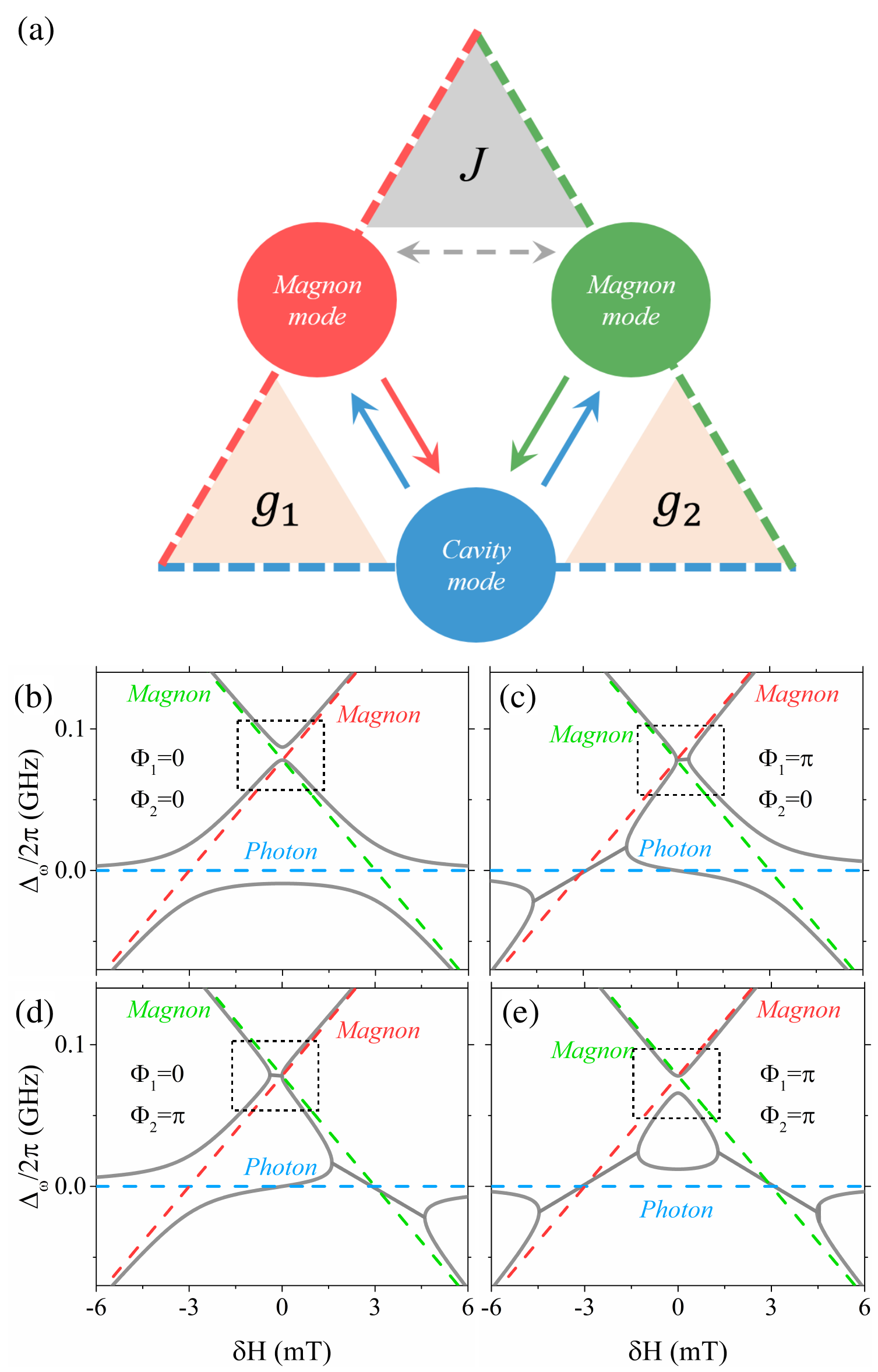,width=8.5cm}
\caption{(a) (Color online) Cavity mediated coupling between two magnon modes. Combining the effect of Faraday's law (indicated by red and green arrows), the competition of Ampere's law and the cavity Lenz effect (indicated by blue arrow) produces coherent or dissipative coupling between the individual magnon mode and the cavity mode with a coupling strength of $g_{1,2}$. By exchange of virtual photons, the two magnon modes are strongly coupled with an exchange coupling rate of $J$(grey dashed arrow). (b)-(e) The hybridized mode frequency $\Delta_\omega=\omega-\omega_c$ (solid gray lines) is plotted as a function of the local magnetic field $\delta H$, which is used to control the frequency difference of $\omega_{m1,2}$. The red and green dashed lines represent uncoupled magnon modes while the blue dashed line represents cavity mode. The coupling effects of the system produce level repulsion and level attraction of two magnon modes indicated by dotted box. During the calculation, we set $g_1/2\pi=g_2/2\pi$=20 MHz and the magnetocrystalline anisotropy field $H_{A1,A2}$=0.}
\label{Fig2}
\end{center}
\end{figure}

Based on the key features of level repulsion and attraction shown in Sec. \ref{Sec:2x2model}, now we study a three-mode coupled magnon-photon system, where two spatially separated magnon modes, $m_1$ and $m_2$, couple with a cavity mode ($a$). The schematic diagram of this three-mode coupled system is illustrated in Fig. \ref{Fig2}(a). The microwave current drives or impedes the dynamics of magnetization through the competition of Ampere's law and the cavity Lenz effect (indicated by blue arrow).\cite{Harder2018PRL} Meanwhile, due to the effects of Faraday's law, the magnetization precession also creates a back action effect onto the cavity field (indicated by red and green arrows). Thus the two magnon modes are coupled to the cavity mode with a coupling strength of $g_{1,2}$. By exchanging virtual photons, the two magnon modes are strongly coupled with an exchange coupling rate of $J$(grey dashed arrow).

The equivalent non-Hermitian Hamiltonian of this three-mode system can be written as

\begin{eqnarray}
H=&\hbar \omega_ca^\dag a+\hbar\omega_{r1} m_1^\dag m_1+\hbar g_1(a^\dag m_1+e^{i\Phi_1}am_1^\dag)\\ \nonumber
&\hbar\omega_{r2} m_2^\dag m_2+\hbar g_2(a^\dag m_2+e^{i\Phi_2}am_2^\dag),
\label{Eq:3x3Hamiltonian}
\end{eqnarray}

\noindent where $\omega_{m1,m2}$, $g_{1,2}$ and $\Phi_{1,2}$ are the frequency, coupling strength to the cavity mode and coupling phase for the magnon mode $m_{1,2}$, respectively. Assuming sufficient spatial separation between the two magnetic samples, we neglect direct interactions between $m_1$ and $m_2$. \cite{Lambert2016PRA} The hybridized eigenmodes of the system can be solved from the equation of motion  

\begin{equation}
\frac{d}{dt}
\begin{pmatrix}
a\\ m_1\\ m_2
\end{pmatrix}=
i\begin{pmatrix}
\omega_c & g_1 & g_2\\
e^{i\Phi_1}g_1 & \omega_{m1} & 0 \\
e^{i\Phi_2}g_2 & 0 &\omega_{m2}
\end{pmatrix}
\begin{pmatrix}
a\\ m_1\\ m_2
\end{pmatrix}.
\label{Eq:3x3matrix}
\end{equation}

For this three-mode coupled system with two identical YIG spheres placed within a cavity, a global magnetic field $H$ is applied to tune the frequency of the magnon modes according to $\omega_{m1,m2}(H)=\gamma(H+H_{A1,A2})$, where $\gamma=2\pi\times27.6$ GHz/T is gyromagnetic ratio, and $H_{A1,A2}$ is the anisotropy field for each individual YIG sample. In the meantime, the field at each sphere can be locally adjusted by $\pm\delta H/2$ via a small coil. In the following discussion, we operate the system in the dispersive limit, where both magnons are significantly detuned from the cavity ($|\Delta_{1,2}|$=$|\omega_{m1,m2}-\omega_c|\gg{}g_{1,2}$).

Figures 2(b-e) show the calculated dispersions of the hybridized modes for four cases based on the coupling state of $(\Phi_1, \Phi_2)$, describing either coherent or dissipative coupling between the individual magnon modes and the cavity mode. Focusing on the region highlighted by a dotted box, where the two magnon modes (red and green dashed lines) cross each other, level repulsion is observed for (0, 0) and ($\pi$, $\pi$) states, while level attraction is observed for (0, $\pi$) and ($\pi$, 0) states. In order to give a more detailed explanation for this striking feature, we analytically solved Eq. (\ref{Eq:3x3matrix}) in the dispersive limit and rewrite the frequencies of hybridized modes in an explicit form, which is similar to Eq. (\ref{Eq:2x2eigenfrequency}) for directly coupled $a\!-m$ system, as

\begin{equation}
\omega_{m\pm}=\frac{1}{2}\left[\omega'_{m1}+\omega'_{m2}\pm\sqrt{(\omega'_{m1}-\omega'_{m2})^2
+4e^{i(\Phi_1+\Phi_2)}J^2}\right],
\label{Eq:two-magnon-eigenfrequency}
\end{equation}

\noindent where $\omega'_{m1,m2}=\omega_{m1,m2}+e^{i\Phi_{1,2}}g_{1,2}^2/\Delta_{1,2}$ includes a finite Lamb shift \cite{Majer2007Nature, Blais2007PRA, Grigoryan2019} of the energy level, which can be either blue or red shift dependent on not only the sign of detuning $\Delta_{1,2}$ but also the nature of the coupling between the magnon and cavity. $J=\frac{1}{2}g_1g_2|\frac{1}{\Delta_1}+\frac{1}{\Delta_2}|$ is the effective coupling strength between $m_1$ and $m_2$. 

Equation (\ref{Eq:two-magnon-eigenfrequency}) indicates that the indirect coupling features of the long-range magnon-magnon interaction are solely determined by the phase between them ($\Phi_1$+$\Phi_2$): level repulsion for $\cos(\Phi_1$+$\Phi_2)=1$ and level attraction of $\cos(\Phi_1$+$\Phi_2)=-1$. These relations well explains the coupling signature highlighted in Fig. 2(b-e).

Fig. 3(a) and (b) show a zoomed-in view of the boxed areas in Fig. 2(b) and (d). A careful examination shows that the hybridization modes always cross $\omega_{m1}$ at the point where $\omega_{m1}=\omega_{m2}$, which is different from the observations in directly coupled system [Figs. 1(a) and (b)]. Mathematically, this point results from the Lamb shift, which causes a shift in the frequency detuning $\Delta_\omega$ with a magnitude of $J$ for level repulsion and a shift in field detuning $\Delta_H$ with a magnitude of $2J$ for level attraction. Physically, this point is related to "dark" states of coupled systems \cite{Majer2007Nature, zhang2015NC} where the hybridization of two magnon modes precess out of phase with an identical amplitude, and as a consequence, their interactions are decoupled from the cavity mode. 

We can understand this effect by deducing the eigenvector of the $m_1\!-\!m_2$ subsystem

\begin{eqnarray}
\begin{pmatrix} m_1\\ m_2 \end{pmatrix}
&=\begin{pmatrix}  e^{i\Phi_1}J \\
  \delta/2\pm\sqrt{\delta^2/4+e^{i(\Phi_1+\Phi_2)}J^2} \end{pmatrix},\nonumber \\
&\mathrm{with}\qquad \delta= \omega'_{m1}-\omega'_{m2}. 
\label{Eq:two-magnon-eigenvector}
\end{eqnarray}

Following the similarity between Eqs. (\ref{Eq:two-magnon-eigenvector}) and (\ref{Eq:2x2eigenvector}), the phase correlation $\phi_{m2}-\phi_{m1}$ can be determined exactly the same way as $\phi_{m}-\phi_{a}$ for the directly coupled $a\!-\!m$ system shown in Fig. 1(c) and (d).  Using the approximation of $\Delta_1\simeq\Delta_2\simeq\Delta$ in the dispersive limit, the photon part of the eigenstate can be deduced as

\begin{equation}
a=\frac{g_1m_1+g_2m_2}{\Delta}.
\label{Eq: 3x3 cavity mode}
\end{equation}

Combining Eqs. (\ref{Eq:two-magnon-eigenvector}) and (\ref{Eq: 3x3 cavity mode}), we can determine that the dark state ($a=0$) appears at 
\begin{equation}
\delta=\frac{e^{i\Phi_1}g_1^2-e^{i\Phi_2}g_2^2}{\Delta}.
\label{Eq: detuning of dark mode}
\end{equation}

\begin{figure} [t]
\begin{center}\
\epsfig{file=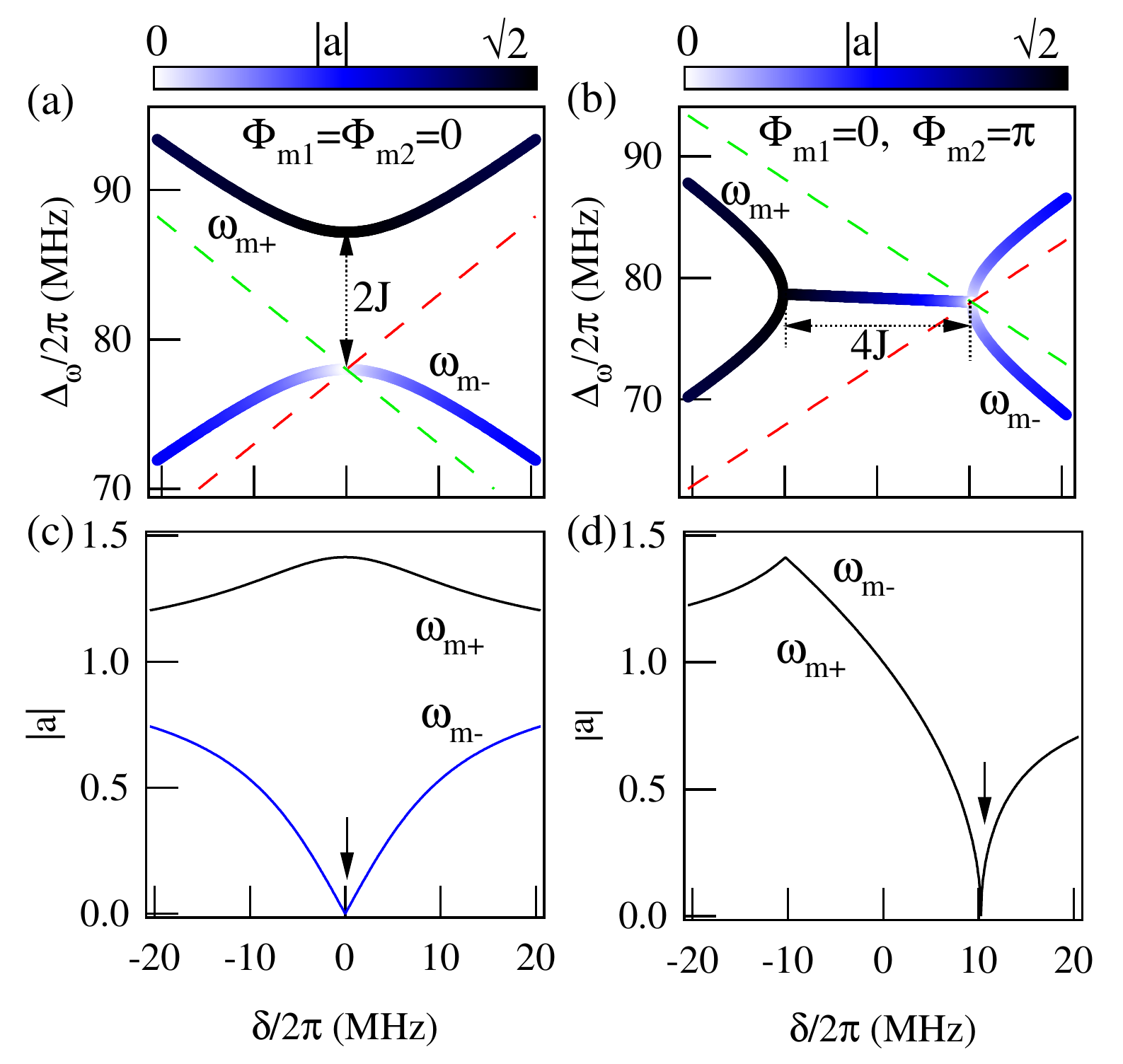,width=8.5cm}
\caption{(a),(b) The hybridized dispersions of indirect long-range interacted magnon modes for level repulsion and attraction calculated based on Eq. (\ref{Eq:two-magnon-eigenfrequency}), respectively, which are coloured by the magnitude of $|a|$.  The parameters are $g_1/2\pi=g_2/2\pi=g/2\pi$=20 MHz and $\Delta/2\pi$=78 MHz. The dashed lines indicate the uncoupled dispersion of the magnon modes. (c), (d) The calculated $|a|$ is based on Eq. (\ref{Eq: 3x3 cavity mode}) for level repulsion and level attraction, respectively, normalized by a factor $g/\sqrt{|m_1|^2+|m_2|^2}$ for clarity. Arrows indicate the dark state.}
\label{Fig3}
\end{center}
\end{figure}

As an example, the calculated $|a|$ (normalized by a factor $g/\sqrt{|m_1|^2+|m_2|^2}$ for clarity) is plotted in Figs. 3(c) and (d) for $g_1$=$g_2$ and $\Delta>0$, where the dark state clearly appears at $\delta=0$ on the $\omega_{m-}$ branch for $\Phi_1=\Phi_2=0$ (level repulsion), and at $\delta=2J$ on the right exceptional point for $\Phi_1=0$ and $\Phi_2=\pi$ (level attraction). Since $\phi_{m2}-\phi_{m1}$ for $\omega_{m+}$ and $\omega_{m-}$ only differ in sign for the level attraction case [similar to $\phi_m-\phi_a$ in Fig. 1(d)], it does not affect the amplitude of $g_1m_1+g_2m_2$ and hence $|a|$. As a result, $|a|$ is identical for both $\omega_{m+}$ and $\omega_{m-}$ branches as shown in Fig. 3(d).  

In previous studies\cite{FilippPRA2011}, the dark state always occurs at the hybridized mode closer to the cavity mode in frequency. Here, we find that the dark state of long-range coherent coupling may also reside in the outer branch of the hybridized modes if both magnon modes are dissipatively coupled to the cavity mode ($\Phi_1=\Phi_2=\pi$). Our model indicates that the dark state can be adjusted by phases $\Phi_1$ and $\Phi_2$ in addition to the sign of $\Delta$. \cite{Lambert2016PRA}

If $g_1\neq{}g_2$, the dark state appears away from these symmetric points on the hybridized magnon-magnon dispersion. However, by substituting the lamb shift into Eq. (\ref{Eq: 3x3 cavity mode}) one can find a general relation for the dark state where $\omega_{m1}=\omega_{m2}$, regardless of the detailed coupling feature between the individual magnons and the microwave cavity. Here, the dark mode is induced by the hybridization of the two magnon modes when they precess out of phase,  $ \begin{pmatrix} m_1\\m_2 \end{pmatrix} = \begin{pmatrix} -e^{i\Phi_1}J\\e^{i\Phi_1}g_1^2/\Delta \end{pmatrix} $, and their coupling effects on the cavity mode cancel each other, resulting a vanishing total response of the magnon dynamics to the cavity mode.
 
\section{Experiment results and discussion}

\begin{figure} [t]
\begin{center}\
\epsfig{file=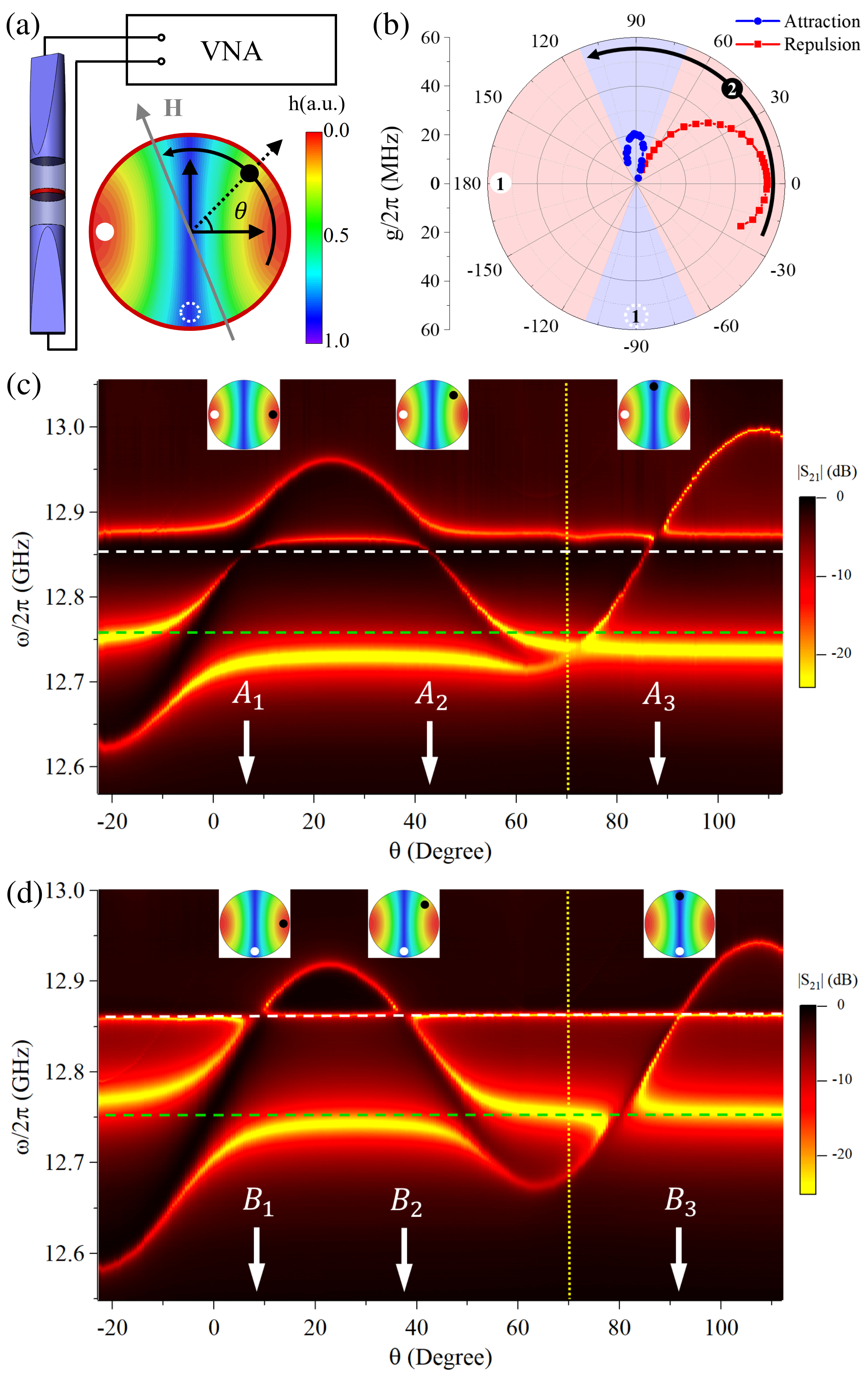,width=8.4 cm}
\caption{(a) Experimental setup, with a VNA measuring the microwave transmission through a waveguide loaded with two YIG spheres. The simulated $h$ field amplitude for the cavity mode at the middle plane. YIG1 (white) is fixed at either a node or antinode of the $h-$ field, while we rotate YIG2 (black) anticlockwise covering areas of coherent and dissipative coupling. The static bias field $H$ is applied along $\theta\simeq$112$^\circ$. (b) The net coupling strength $g_2$ as a function of YIG2 position $\theta$. The red and blue areas represent different regimes of coherent coupling and dissipative coupling, respectively. Experimental transmission spectra as a function of position angle of YIG2 when YIG1 is located at the $h-$ (c) node or (d) antinode. The yellow dotted line stands for 70.5$^\circ$ where the coupling regime switches. The dashed lines indicate the uncoupled modes of $\omega_{m1}$ and $\omega_c$ as a guide to the eye. The insets show the approximate positions, $A_i$ and $B_i$ of the YIG spheres when they strongly interact.}
\label{Fig4}
\end{center}
\end{figure}

The experimental setup of our measurement system is schematically shown in Fig. 4 (a). The microwave cavity used in this work is a Fabry-Perot-like cavity based on the Ku band (12-18 GHz) assembled waveguide apparatus, where circular waveguides are connected through circular-rectangular transitions to coaxial-rectangular adapters, and the two transitions are rotated by an angle of 45$^\circ$. \cite{Yao2015PRB} The inner diameter of the circular waveguide is 16.1 mm. The indirect magnon-magnon coupling is studied by placing two identical 1-mm diameter single crystal YIG spheres in the midplane of this quasi-one-dimensional cavity.  Both YIG sphere is placed approximately 2 mm from the inner edge of the waveguide.

For our cavity resonance we use the TE$_{11}$ mode (where the overall electric field is maximum at the midplane of our cavity) at $\omega_c/2\pi$=12.76 GHz.\cite{Yao2015PRB} The intrinsic damping parameters are $\alpha=7.60 \times10^{-5}$ and $\beta=8.49 \times10^{-3}$ for the magnon and cavity modes, respectively. During the measurement, a constant static magnetic field, $H$, is applied along $\theta=112^\circ$ ($\theta$ is defined as $0^\circ$ or $180^\circ$ for $h-$ antinodes). Two samples labeled as YIG1 and YIG2 are carefully mounted on the fixed and rotatable part of a waveguide insert. The special design enables us to rotate YIG2 around the cavity axis within a angular precision of 0.5$^\circ$. The profile of the microwave magnetic (h-) field the midplane was simulated using Computer Simulation Technology Microwave studio, which is shown in Fig. 4(a). When changing the position of YIG2 by rotating the waveguide insert, $g_2e^{i\Phi_2}$ evolves dramatically from coherent coupling to dissipative coupling with $\theta$. \cite{Harder2018PRL} Meanwhile, the direction of the magnetocrystalline anisotropy field of YIG2 also rotates relative to the external field, producing an oscillating local field of $H+H_{A2}$, and a sinusoidal dispersion of $\omega_{m2}$. \cite{McKinstryJAP1985, ArtmanPR1957} By fixing one YIG sphere and rotating the other, we are able to precisely manipulate the frequency detuning $\omega_{m1}-\omega_{m2}$ as well as the coupling regime. Using a vector network analyzer (VNA) we measure the microwave transmission $S_{21}$ of this three-mode system.

We first calibrate the coupling effects between the cavity mode and a single YIG sphere. YIG1 is fixed at a position with an angle either $\theta=180^\circ$ ($h-$ antinode) or $\theta=-90^\circ$ ($h-$ node). The S-parameter measurement for this single YIG allow us to determine $g_1/2\pi$=55 MHz and $\Phi_1=0$ for $\theta=180^\circ$ and $g_1/2\pi$=19 MHz and $\Phi_1=\pi$ for $\theta=-90^\circ$. Separately, YIG2 is rotated over an angle range of $-22^\circ\leqslant\theta\leqslant112^\circ$ covering two distinct coupling regimes of coherent coupling and dissipative coupling. The deduced coupling strength, $g_2$, as a function of the angular position $\theta$ for YIG2 are summarized in Fig. 2(b), from which we found that the critical angle where the coupling regime switches is 70.5$^\circ$.

By placing YIG1 at $\theta=180^\circ$ corresponding to the coherent coupling region of $g_1$ and rotating YIG2 within the cavity, we measure the long-range coupling between magnon modes mediated by the cavity photon. Figure 4(c) shows the results for a fixed static magnetic field $\mu_0H$=488 mT, where $\omega_c/2\pi$=12.76 GHz and $\omega_{m1}/2\pi$=12.85 GHz indicated by dashed lines are the uncoupled cavity mode and YIG1 magnon mode frequencies. By rotating YIG2, the interaction between the two magnon modes is clearly seen in Fig. 4(c) when $\omega_{m2}$ approaches $\omega_{m1}$, indicated by arrows label as $A_{1-3}$. At conditions $A_1$ and $A_2$, where both $\omega_{m1}$ and $\omega_{m2}$ are coherently coupled with $\omega_c$, the long-range interaction between the two magnon modes shows a characteristic feature of the avoided level crossing. As we rotate YIG2 clockwise across 70.5$^\circ$, the two spheres enter different coupling regions. When their frequencies again meet at the condition $A_3$, level attraction between the indirectly coupled magnon modes is experimentally demonstrated for the first time.  

Next we place YIG1 at $\theta=-90^\circ$ corresponding to the dissipative coupling region of $g_1$ and repeat the above measurement. The results are summarized in Fig. 4(d). Despite the same rotation trajectory of YIG2, the behavior of the long-range interaction between the magnon modes shows a different pattern. At conditions $B_1$ and $B_2$, where the two YIG spheres have different coupling mechanisms to the cavity mode, a characteristic feature of level attraction occurs. Meanwhile, level repulsion is observed when both YIG spheres are dissipatively coupled with the cavity. 

This experiment unambiguously validates our model, demonstrating long-range indirect coherent and dissipative interactions between two spatially separated magnons.  The characteristic features of the long-range interaction are solely determined by the relative phase between $\Phi_1$ and $\Phi_2$.  Furthermore, the dark state is clearly seen at the condition $A_1$, where $g_1\simeq{g_2}$. At conditions $A_2$, $A_3$, $B_1$ and $B_2$, the dark state predicated by Eq. (\ref{Eq: detuning of dark mode}) appears far away from the strongly coupling regime due to the significant difference between $g_1^2$ and $g_2^2$ and is thus not well resolved in current experiment.

To quantitatively explain the experimental observation of long-range interactions between two magnon modes, we calculated the dispersion by solving the determinant of Eq.(\ref{Eq:3x3matrix}). We focus on the three conditions $A_1$, $B_1$, and $B_3$, which correspond to three typical cases of long-range magnon-magnon interaction: (a) both magnon modes coherently coupled with the cavity, (b) one magnon mode coherently and the other dissipatively coupled with the cavity, and (c) both magnon modes dissipatively coupled with the cavity. For the calculation, $g_1$ and $g_2$ are determined by experimental measurements. For simplicity, we assume $\omega_{m2}$ follows a relation as $\omega_{m2}-\omega_{m1}\propto\sin(\theta-\theta_0)+C$ within a 20$^\circ$ range where $\theta_0$ and $C$ are constants. As shown in Fig. 5, the comparison between simulation and experiment illustrates a quantitative agreement. 

\begin{figure} [t]
\begin{center}\
\epsfig{file=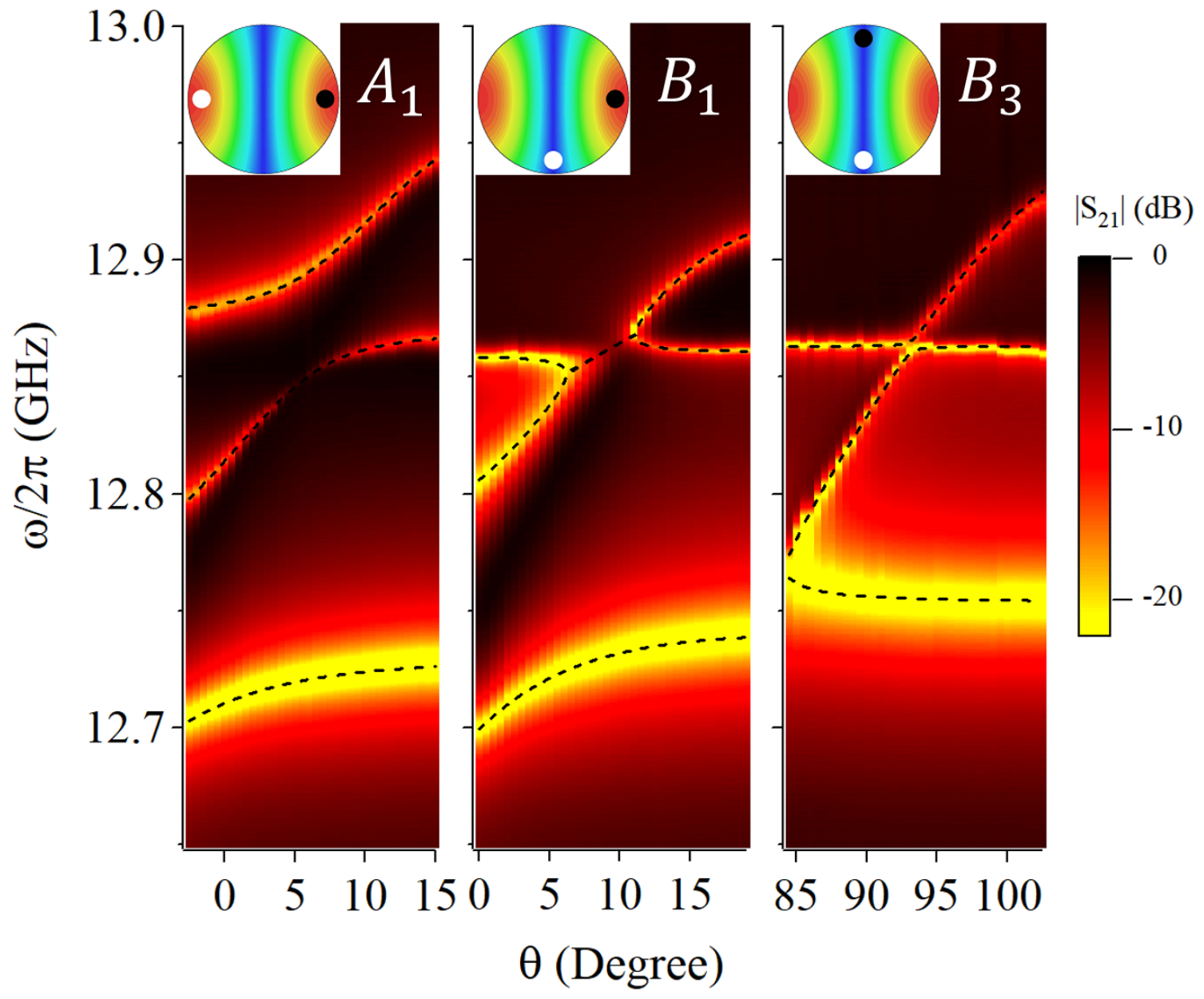,width=8.5 cm}
\caption{Zoomed-in view of transmission spectra at conditions $A_1$, $B_1$ and $B_3$ in Fig.4(c) and (d), corresponding to three coupling states of ($\Phi_1$, $\Phi_2$), i.e. (0,0), (0, $\pi$) and ($\pi$, $\pi$). The black dashed lines indicate calculated dispersion by solving the determinant of Eq.(\ref{Eq:3x3matrix}) using the measured coupling strength $g_1/2\pi$= 55 MHz and $g_2/2\pi$=52 MHz for $A_1$, $g_1/2\pi$= 19 MHz and $g_2/2\pi$=52 MHz for $B_1$, and $g_1/2\pi$= 19 MHz and $g_2/2\pi$=20 MHz for $B_3$. }
\label{Fig5}
\end{center}
\end{figure}

Thanks to the high sensitivity of our experimental implementation, we can study the dependence of the indirect coupling strength $J$ on detuning $\Delta$ by varying the static magnetic field $H$. As clearly seen in Fig. 6(a), the amplitude of the hybridized magnon modes decreases with increasing $\Delta$, and furthermore the gap between the hybridized magnon modes shrinks. Although amplitude of the hybridized magnon modes gradually decrease, the dark state is well resolved. 

For the case of indirect coherent coupling, the coupling strength $J$ can be directly determined from the polariton gap (=$2J$) of the dispersion as indicated in Fig. 3(a). The measured amplitude of $J$ is plotted as black squares in Fig 6(b). To compare with our model, we first calculate the dispersion [dashed lines in Fig. 6(a)] using Eq. (\ref{Eq:3x3matrix}), which is in agreement with experimental results. During the calculation one set of parameters ($g_1/2\pi=$55 MHz, $g_2/2\pi=$52 MHz, and $\Phi_1=\Phi_2=0$) was used. The $J$ deduced from our calculations (solid line) is in agreement with experimental results. In the dispersive limit, the predicted value (dashed line) follows $J=g_1g_2/\Delta$. Comparing this to the experimental results, it is overestimated in the strong coupling range. In the strongest coupled case at $\omega_{m1}=\omega_{m2}=\omega_{c}$, the gap between two hybridized magnon modes is $2\sqrt{g_1^2+g_2^2}$ rather than infinite value at $\Delta=0$ predicted for the dispersive limit.

A similar dependence is also revealed for long-range indirect dissipative coupling in Fig. 7. Here the coupling strength $J$ is directly determined from the range (=$4J$) of the coalescent of hybridized magnon modes[as indicated in Fig. 3(a)]. The calculated dispersion [dashed lines in Fig. 7(a)] used $g_1/2\pi=$19 MHz, $g_2/2\pi=$20 MHz, $\Phi_1=\pi$, and $\Phi_2=0$ determined by independent experiments. Again, when $\Delta$ becomes comparable to $g_i$, all three modes are highly hybridized, leading to the break down of the dispersive approximation. 

\begin{figure} [t]
\begin{center}\
\epsfig{file=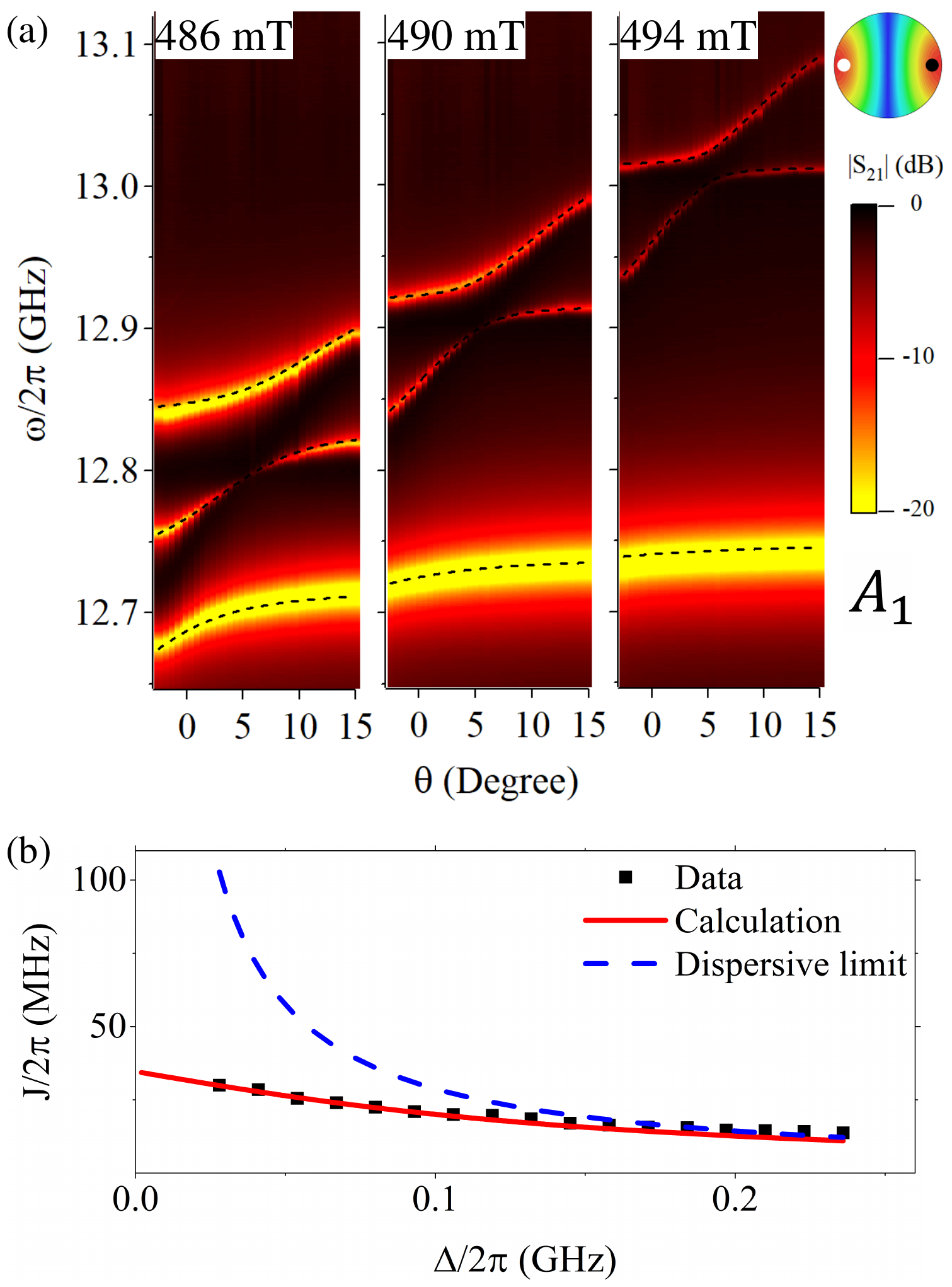,width=8.5 cm}
\caption{(a) For long-range coherent coupling, transmission spectra at various external field demonstrate the decay of the coupling strength $J$ when increasing $\Delta$. (b) $J$ as a function of $\Delta$ at $A_1$ when both magnon modes are coherently coupled with the cavity mode. Black squares are determined by measured data in (a), while the red line represents calculation result based on Eq.(\ref{Eq:3x3matrix}) and the blue dashed line follows the $g_1g_2/\Delta$ dependence in the dispersive limit. }
\label{Fig6}
\end{center}
\end{figure}

\begin{figure} [t]
\begin{center}\
\epsfig{file=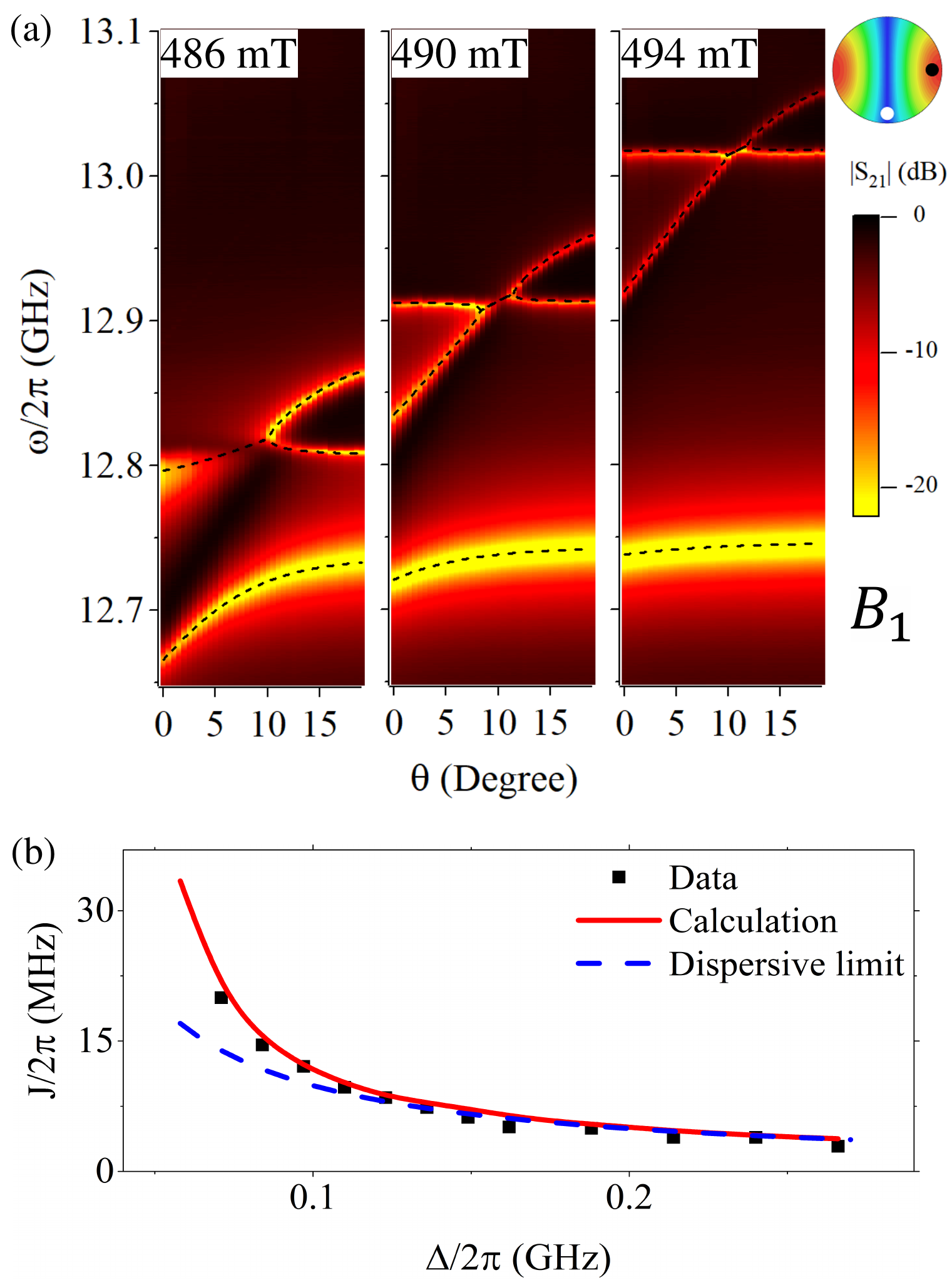,width=8.5 cm}
\caption{(a)For long-range dissipative coupling, transmission spectra at various external field demonstrate the decay of the coupling strength $J$ with increased $\Delta$. (b) $J$ as a function of $\Delta$ at $B_1$ with one magnon coherently and one dissipatively coupled with the cavity mode. Black squares are determined by measured data in (a), while the red line represents the calculation results based on Eq.(\ref{Eq:3x3matrix}) and the blue dashed line follows $g_1g_2/\Delta$ dependence in the dispersive limit. }
\label{Fig6}
\end{center}
\end{figure}

\section{Conclusions}
We have presented a systematic study of the effect of indirect coupling between two magnon modes mediated by a cavity mode. A theoretical model based on a phenomenological approach was developed to describe the dispersions and phase information of the system in the dispersive limit. The characteristic properties of XOR-like coupling relations and magnon dark states are revealed both theoretically and experimentally. Putting magnets on a similar basis to qubits and atoms in cavities, our work provides a new method for studying cavity mediated coupling in the framework of cQED. Furthermore, in a general context, our model system demonstrates the transition rule of the coupling state where two sub-systems interact with each other through a mediating oscillator, which can act as a building block to further understand long range light-matter interactions.

\section*{Acknowledgements}

This work was funded by NSERC Discovery Grants and NSERC Discovery Accelerator Supplements (C.-M. H.). P.-C. Xu was supported in part by China Scholarship Council. We thank Y. Zhao for helpful discussions and suggestions. We also thank P. Hyde, Y. Wang for proofreading the manuscript. 

Note added.-During the manuscript preparation, we found a preprint theoretical work on dissipative long-range magnon-magnon interactions in a coupled magnon-photon system[Ref. \onlinecite{Grigoryan2019}], which presents an alternative consistent theoretical picture for our experiment.

\end{document}